# Defect States in Hexagonal Boron Nitride: Assignments of Observed Properties and Prediction of Properties relevant to Quantum Computation.

A. Sajid,[1,2] Jeffrey R. Reimers,[1,3] and Michael J. Ford[1*]

[1]School of Mathematical and Physical Sciences, University of Technology Sydney, Ultimo, New South Wales 2007, Australia
[2]Department of Physics, GC University Faisalabad, Allama Iqbal Road, 38000 Faisalabad, Pakistan
[3]International Centre for Quantum and Molecular Structures and School of Physics, Shanghai University, Shanghai 200444, China

*Mike.Ford@uts.edu.au, Jeffrey.Reimers@uts.edu.au

**Abstract:**

Key properties of 9 possible defect sites in hexagonal boron nitride (h-BN), $V_N$, $V_N^{-1}$, $C_N$, $V_NO_{2B}$, $V_NN_B$, $V_NC_B$, $V_BC_N$, $V_BC_NSi_N$, and $V_NC_BSi_B$, are predicted using density-functional theory (DFT) that are corrected by applying results from high-level *ab initio* calculations. Observed h-BN electron-paramagnetic resonance (EPR) signals at 22.4 MHz, 20.83 MHz, and 352.70 MHz are assigned to $V_N$, $C_N$, and $V_NO_{2B}$, respectively, while the observed photoemission at 1.95 eV is assigned to $V_NC_B$. Detailed consideration of the available excited states, allowed spin-orbit couplings, zero-field splitting, and optical transitions is made for the two related defects $V_NC_B$ and $V_BC_N$. $V_NC_B$ is proposed for realizing long-lived quantum memory in h-BN. $V_BC_N$ is predicted to have a triplet ground state, implying that spin-initialization by optical means is feasible and suitable optical excitations are identified, making this defect of interest for possible quantum-qubit operations.

# I.    INTRODUCTION

Point defects in semiconductors show a rich spin and optoelectronic physics that can be exploited to fabricate qubits for quantum computing technology [1,2] as well as for single photon sources for quantum cryptography. Currently, the negatively charged nitrogen vacancy centre ($N_v^{-1}$) in diamond is the leading candidate not only as a single photon source for quantum cryptography [3,4], but also as an optically coupled quantum register for scalable quantum information processing. Possible applications include those in both quantum communication [5] and in distributed quantum computation [6]. However, useful qubits can also be conceived based on other semiconducting materials. Hexagonal boron nitride (h-BN) is a wide band gap (~6 eV) 2-dimensional (2-D) material with the potential to host many such colour centres [7-14] that are promising candidates for quantum applications.

Both unintentional occurrence of defects (e.g. vacancies) [15,16] during the preparation of single layer h-BN and the deterministic production of defects through electron beam radiation [17] have been shown to have considerable effects on electronic and magnetic properties of h-BN. While both nitrogen ($N_v$) and boron ($B_v$) vacancies can act as paramagnetic centres in h-BN [18], electron paramagnetic resonance (EPR) studies indicate that $N_v$ are more important [19-24]. Two types of paramagnetic centres have been identified: (i) three-boron centres (TBC) in which an unpaired electron interacts with three equivalent boron ($B^{11}$) nuclei, producing 10-line EPR spectra, and (ii) one-boron centres (OBC) in which oxidative damage at the centre forces the unpaired electron to interact with only a single $B^{11}$, producing 4-line EPR spectra. The TBC can be deliberately produced either by irradiation [19-21] or by carbon doping [23], but controllable h-BN oxidation to produce OBCs has not yet been achieved. Of particular interest is that carbon-doped centres can give rise to intense photoluminescence (PL) [25,26]. Although speculations based on EPR studies have been made about the form of the TBC and OBC centres [19-24], their exact nature remains uncertain.

A-priori calculations using density-functional theory (DFT) can provide useful tools for the interpretation of EPR data such as observed hyperfine tensors. Indeed, such calculations have been helpful in the identification of point defects in different semiconductors by comparing the experimental and calculated hyperfine constants [27-31]. However, the nature of many ground or excited states relevant to defects and their applications are such that DFT methods deliver results of widely varying accuracy [32], meaning that great care must always be taken when this method is applied. A summary of results comparing DFT calculations to high-level ab initio calculations for a model defect in h-BN, $V_NC_B$, concludes that DFT is reliable for the triplet-state manifold but *underestimates* the stability of closed-shell singlet states by around 0.5 eV and *overestimates* the stability of open-shell singlet states by around 0.5 eV [32], meaning that open-shell singlet states are predicted to be too low in energy by of order 1.0 eV compared to closed-shell singlet states. In the context of this work, these errors indicate that DFT should properly describe the magnetic properties of states but the nature of the ground state may not be correctly predicted. Further, we have also considered the effects of zero-point energy correction and free-energy corrections at 298 K [32], finding sizeable changes in relative state energies of up to 0.25 eV, but corrections of this magnitude remain small compared to effects of interest herein and so are neglected.

To the best of our knowledge, there is no comprehensive theoretical study on the hyperfine tensors of defects in h-BN and the prospects for their exploitation in quantum computation. We consider detailed models of the TBC and OBC defects in h-BN, as well as proposing many new defect centres, particularly defects involving introduced carbon impurity atoms. Key known properties of the TBC and OBC defects are reproduced including details of their doublet ground states. Various carbon-related anti-site defects are also considered in which a nitrogen or boron vacancy is accompanied by a neighbouring carbon-atom and/or silicon-atom substitution. In particular, defects with possible ground state conformers of triplet character are searched for, a desirable feature exploitable for quantum computation. Group-theoretical analysis is used to guide how such applications could develop, suggesting new directions for experimental studies. These results are combined with our recent analyses of PL properties of h-BN defects [33] to allow focus on possible quantum qubit applications.

## II. COMPUTATIONAL METHODS

Calculations are performed for periodically replicated defects in 2-D h-BN nanoflakes using DFT. For calculation of total energy, electronic structure and ground state geometry of all the defect supercells we used version 5.3.3 of the Vienna Ab Initio Simulation Package (VASP) [34,35]. For accurate calculation of electron spin density close to the nuclei, the projector augmented wave method (PAW) [36,37], was applied together with a plane wave basis set. We utilized the standard PAW-projectors provided by the VASP package. Pristine single-layer h-BN was first geometrically optimized using the conventional cell and a 21×21×1 Monkhurst-Pack reciprocal space grid. A large vacuum region of 30 Å width was used to separate a single layer of h-BN from its periodic images and to ensure that interaction between periodic images is negligible. The optimized bond length of pristine h-BN is 1.452 Å. All the defects were then realized in a 7x7x1 supercell and allowed to fully relax using a plane wave cut-off of 350eV for a maximum force of 0.01eVÅ$^{-1}$. These optimized supercells were then embedded in 10x10x1 supercells using a similar method as described in Ref. [38] for minimizing the computational cost for large calculations and to ensure no interaction in planer directions for the calculation of hyperfine coupling(HF) constants. An increased plane-wave cut-off of 500 eV was used, which was sufficient to obtain a converged spin density and HF coupling constants ($A_{xx}$, $A_{yy}$, $A_{zz}$), where $A_{xx}$, $A_{yy}$, $A_{zz}$ are the principle values of HF coupling tensor. For these supercells, the Γ-point sampling of the first Brillouin zone sufficed. We have used the non-local Heyd-Scuseria-Ernzerhof hybrid functional (HSE06) [39,40] by taking into account the contribution of the spin polarization of the core electrons to the Fermi-contact term. This has been shown to produce accurate results for HF tensor calculation for point defects in semiconductors [28]. Zero-field splitting tensors are evaluated using the method described in Ref. [41]. For the calculation of the spin-spin contribution to the zero-field tensor, a higher cut-off energy of 600 eV and lower force tolerance of $10^{-4}$ eVÅ$^{-1}$ were used. Optimized geometries from VASP were used to calculate the wave function coefficients of defect states using Siesta [42,43].

The band gap of pristine h-BN is calculated to be 5.69eV, which is in reasonable agreement with the experimental value [44]. It is further seen that by changing the alpha parameter (from 0.25 to 0.75) in the HSE functional, the value of band gap for pristine h-BN increases linearly (from 5.69 to 7.85eV). Since the exact reproduction of experimental gap by adjustment of alpha parameter does not guarantee the accurate prediction of defect levels, in the present study we stick with the standard value of alpha parameter of 0.25 for HSE06. The accuracy of defect levels is ensured by calibrating the individual levels against *ab initio* CCSD(T), EOMCCSD, CASPT2, and MRCI calculations for a model compound [32].

The energies of open-shell singlet states are calculated using spin relaxed calculations ("ISPIN=2") with equal numbers of spin-up and spin-down electrons ("NUPDOWN=0") using the "FERDO" and "FERWE" commands to set non-Aufbau occupancies akin to one of the degenerate spin components of an open-shell singlet state, assuming that resulting energy is the average of the associated singlet and triplet states [45]. The resulting wavefunctions are spin contaminated and violate the Gunnarsson-Lundqvist theorem [46] so that the associated densities do not provide legitimate solutions to the basic equations of DFT. However, recent calibration of this procedure for one h-BN defect against ab initio and time-dependent DFT (TDDFT) calculations indicates that the relative energies of different open-shell states are realistically predicted but in general these states are predicted to be 1 eV too low in energy compared to closed-shell singlet states [32]. Corrections for this effect are applied using state-specific values for $V_NC_B$ (see later in Fig. 6) and generic ones for $V_BC_N$ (see later in Fig. 7).

### III. RESULTS AND DISCUSSION

In this section we present all the h-BN defects studied in this work and discuss the calculated principal values of hyperfine coupling tensor of various defects to make comparison with the available experimental data. A general feature of the results is that all optimized structures remained planar despite the presence of significant chemical forces favouring non-planar structures in which missing covalent bonds in defect sites are reintroduced [32]. This means that σ-π separability remains a feature of defects in h-BN, simplifying interpretation of calculated electronic and nuclear structures. Key results are summarised in Table I while expanded results are given in Supporting Data[47] Table S.I.

**TABLE I.** The average principle values of the HF coupling tensor $(A_{xx}+A_{yy}+A_{zz})/3$, in MHz, calculated without (with) core contribution, listing the atoms with dominant spin polarization, for various ground-state (GS) defects in h-BN, with as well as the nature of a related feasible photoluminescent (PL) state and its transition energy $h\nu$.

| Defect | Symm. | GS | | PL | | $h\nu$ (eV) | Atoms | $(A_{xx}+A_{yy}+A_{zz})/3$ |
|---|---|---|---|---|---|---|---|---|
| | | Symm. | Occupancy[h] | Symm. | Occupancy[h] | | | |
| $V_N$ | $D_{3h}$ | $^2A_2''$ | $(a_2'')^1(e')^0$ | $^2E'$ | $(a_2'')^0(e')^1$ | 3.15 | $B_1$-$B_3$ | 32(22)[b] |
| $V_N^{-1}$ | $D_{3h}$ | $^1A_1'$ | $(a_1'')^2(a_2'')^2$ | - | - | - | - | - |
| $C_N$ | $D_{3h}$ | $^2A_1''$ | $(a_1'')^1$ | - | - | - | $B_1$-$B_3$ | -17(-18)[d] |
| $V_NO_{2B}$ | $C_s$ | $^2A''$ | $(a'')^2(1a')^2(a'')^1$ | $^2A'$ | $(a'')^2(1a')^1(a'')^2$ | 1.90[f] | $B_1$ | 554(506)[c] |
| $V_NN_B$ | $C_{2v}$ | $^2B_1$ | $(a_1)^2(1b_1)^1(2b_1)^0$ | $^2B_1$ | $(a_1)^2(1b_1)^0(2b_1)^1$ | 2.12[f] | $N_1$ | 66 |
| $V_NC_B$ | $C_{2v}$ | $^1A_1$ | $(2a_1)^2(1b_1)^0(2b_1)^0$ | $^1B_1$ | $(2a_1)^1(1b_1)^1(2b_1)^0$ | 2.08[g] | - | - |
| $^3V_NC_B$[a] | $C_{2v}$ | $^3B_1$ | $(2a_1)^1(1b_1)^1(2b_1)^0$ | $^3B_1$ | $(2a_1)^1(1b_1)^0(2b_1)^1$ | 1.58[g] | C | 473 |
| $V_BC_N$ | $C_{2v}$ | $^3B_2$ | $(a_2)^1(1b_1)^1(2a_1)^0$ | $^3A_2$ | $(a_2)^1(1b_1)^0(2a_1)^1$ | 1.54 | C | 28 |
| $V_BC_NSi_N$ | $C_s$ | $^2A'$ | $(a'')^2(1a')^1(2a')^0$ | $^2A'$ | $(a'')^2(1a')^0(2a')^1$ | 2.03[f] | $B_1$ | 338 |
| $V_NC_BSi_B$ | $C_s$ | $^2A''$ | $(1a')^2(1a'')^1(2a'')^0$ | $^2A''$ | $(1a')^2(1a'')^0(2a'')^1$ | 0.62[f] | Si | 66 |

[a] The ground state is calculated to be $^1A_1$ and here the results presented are for the lowest-energy triplet state, $^3B_1$.
[b] In ref. [24], a signal is observed of magnitude $|A_{xx}+A_{yy}+A_{zz}|/3 = 22.43\pm1.4$ MHz that is attributed to an Nv centre.
[c] In Ref. [19], a signal is observed of magnitude $|A_{xx}+A_{yy}+A_{zz}|/3 = 117.06$ MHz that is attributed to an $N_V^{-1}$ TBC, as well as a signal at 352.70 MHz attributed to an oxygen-containing $V_NO_{2B}$ OBC.
[d] In Ref. [23], a signal is observed of magnitude $|A_{xx}+A_{yy}+A_{zz}|/3 = 20.83$ MHz that is attributed to the B$^{11}$ atoms in a $C_N$ TBC.
[f] From PBE calculations, see ref. [33]; est. maximum difference to HSE06 is 0.5 eV.
[g] HSE06 after correction based on *ab initio* results, see Figs. 6 and 7; observed value [7,48,49] of 1.95 eV.
[h] List of all defect orbitals within the h-BN conduction and valence bands, with their occupancy in the dominant wavefunction configuration.

## A. Three-boron centre Defects

Three-boron centre defects are associated with the loss of a nitrogen atom from a site in h-BN, a site surrounded by three different boron atoms. They are important as they can act as activation, recombination, absorption, or photosensitivity centres [23,25,26,50]. In experiments, they have been observed to display 10-line EPR spectra [19-24] but their precise chemical nature remains unknown. We consider three possibilities: a simple neutral defect named $V_N$ in which the nitrogen atom is just removed; this with an electron trapped at the vacancy site, named $V_N^{-1}$; and this with a carbon atom replacing the nitrogen, named $C_N$.

### 1. Negatively charged nitrogen vacancy $V_N^{-1}$

It has been proposed that the observed EPR 10-line signal at 117.06 MHz originates from a negatively charged nitrogen vacancy $V_N^{-1}$ that has a triplet ground state [19]. However, our and previous calculations [9] predict that $V_N^{-1}$ has $D_{3h}$ symmetry supporting a closed-shell singlet ground state. Table I lists this result, labelling the ground state as $^1A'$ as well as listing the occupancies in the dominant wavefunction configuration of all orbitals found to lie in the band gap between the h-BN valence band (VB) and conduction band (CB). In this case, only two mid-gap orbitals are found in which 4 electrons need be distributed, giving a simple closed-shell structure expressed in terms of them as $(a_1'')^2(a_2'')^2$. All triplet states are predicted to be of much higher energy than the ground state singlet. As DFT calculations of the type we perform have been found to underestimate the stability of closed shell singlet states compared to triplet states [32], this result is likely to be robust. If $V_N^{-1}$ indeed has a singlet ground state then it cannot possibly be the source of the observed EPR signal.

### 2. Uncharged nitrogen vacancy $V_N$

This defect has been previously modelled and many key properties determined [13]. In the optimized ground-state structure shown in Fig. 1(a), the three boron atoms surrounding the vacancy relax towards each other, forming a triangular structure with $D_{3h}$ symmetry; details of the structure are given in Supporting Data[47] Table S.I. The distance between any two of the boron atoms surrounding the vacancy is optimized to 2.21 Å, significantly shorter than the separation of 2.51 Å found between any two N-N or B-B nearest neighbours in pristine h-BN. Such changes are expected as the atoms surrounding the defect strive to rearrange to eliminate dangling bonds, possibly liberating a lot of energy, opposed by constraint forces coming from the surrounding material [32]. Table I shows the two defect orbitals located within the h-BN band gap, with the ground state (GS) having an electronic configuration of $(a_2'')^1(e')^0$ with $^2A_2''$ π-type symmetry and a net spin polarization of 1.0. The calculated spin density is shown in Fig. 1(a), its major contribution coming from the $2p_z$ orbitals of the three boron atoms surrounding the vacancy (≈ 0.25 $\mu_B$ each), with the remaining contributions arise mostly form the B and N atoms in the next coordination shell.

Significantly, the appearance of two defect orbitals in the band gap allows for the possibility of sharp optical absorption and/or emission spectra [33] involving the ground state and the $^2E'$ excited state, of configuration $(a_2'')^0(e')^1$, the emission energy is listed under column PL in Table I. While such transitions are symmetry forbidden, strong vibronic coupling associated with Jahn-Teller distortion could provide useful PL spectra. The calculated PL energy is 3.15 eV as listed in Table I; this value is slightly underestimated as compared with previous GW calculations [51] as GW overestimates the band gap of h-BN. However, the excited state is open shell as it contains one electron distributed amongst the two components of the $e'$ orbital. This is a different type of open-shell character to that considered in our high-level ab initio calculations [32] and hence likely errors in this value are difficult to estimate.

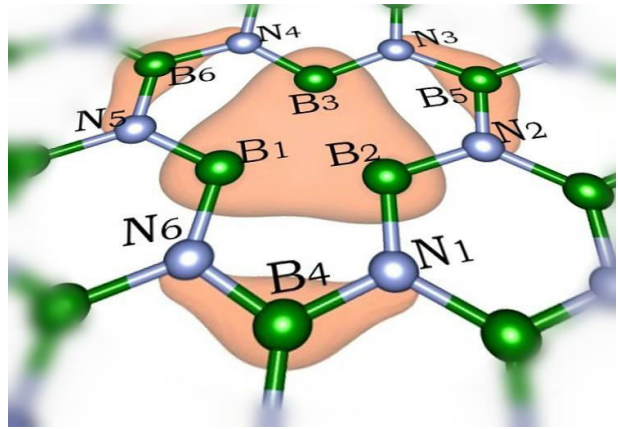

FIG. 1 Geometrical structure and isosurface of the calculated spin density for the neutral $V_N$ centre in h-BN shown from (001) plane at isovalue 0.001/°Å$^3$. Spin density is concentrated on the labelled atoms, providing significant hyperfine couplings.

Supporting Data[47] Table S.I lists calculated HF coupling constants for various atoms surrounding $V_N$. Table I lists the average principle value $(A_{xx}+A_{yy}+A_{zz})/3$ calculated for the HF coupling tensors of the most significant atoms of the defect, $B_1$-$B_3$, as 32 MHz when the calculations exclude core contributions and 22 MHz when this effect is added. An h-BN defect signal has been observed in EPR spectra for the average magnitude of the principle values of the HF tensor (the Fermi-contact term) at 22.43±1.4 MHz and attributed [24] to an Nv type defect. Our calculations support this conclusion and indicate that the signal arises from the three boron-defect atoms numbered $B_1$ - $B_3$ in Fig. 1, as expected.

### 3. $C_N$ centre (substitutional carbon impurity)

The optimized structure for the $C_N$ defect obtained when a carbon atom substitutes a nitrogen atom is shown in Fig. 2. The structure has $D_{3h}$ symmetry and is very similar to pristine h-BN, with neighbouring B-N bond lengths of 1.437Å compared to 1.443 Å before substitution. The ground-state electronic structure is predicted to be $^2A_1''$ with, as previously predicted [9,52] only one orbital in the h-BN band gap. This state has π-type symmetry and net spin-polarization of 1.0. The calculated band gap of a h-BN sheet with a $C_N$ defect of 5.73 eV is in good agreement with previous studies [9,52]. Fig. 2 shows that the unpaired spin localizes prominently on the $p_z$ orbital of the carbon atom. Given the wide range of values that HF coupling constants can take, the calculated average value of the hyperfine coupling constant of -17 MHz is close to the observed magnitude of 20.83 MHz (the sign of HF coupling constant cannot be determined in measurements) [23] in $C^{13}$-enriched carbon doped BN. As only one defect orbital lies in the band gap, sharp PL from this defect is not expected.

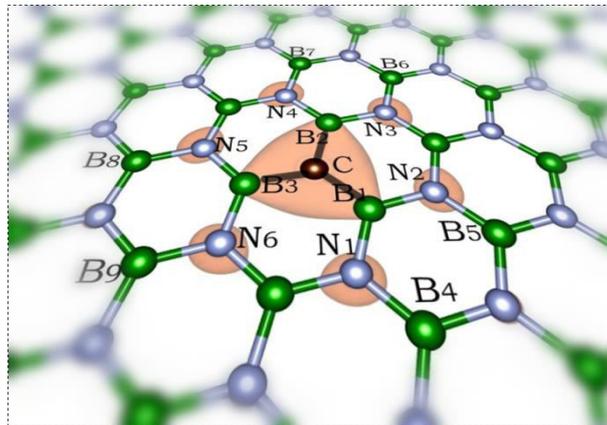

FIG. 2. Geometrical structure and isosurface of the calculated spin density for $C_N$ centre in h-BN shown from (001) plane at isovalue 0.001/°Å$^3$. Spin density is concentrated on the labelled atoms, providing significant hyperfine couplings.

### B. $V_NO_{2B}$ Centre (One-Boron Centre defect)

Oxidation of mono-vacancies in h-BN has previously been studied with the aim of investigating the potential degradation of h-BN in the atmosphere and its impact on electronic and magnetic properties [18]. It has been suggested that oxidized species provides the paramagnetic centre producing observed OBC 4-line EPR spectra [21,53]. To understand this effect, we consider a possible $V_NO_{2B}$ structure in which two oxygen atoms substitute boron at the defect site, making the simplest-possible oxidized OBC. The optimized structure for this defect, which has $C_s$ symmetry, is shown along with its calculated spin density in Fig. 3. In Table I its electronic structure is reported, there being 3 defect orbitals in the h-BN band gap occupied by 5 electrons in the configuration $(a'')^2(1a')^2(2a'')^1$ to make a $^2A''$ ground state. In this, the unpaired electron is concentrated on atom $B_1$ and polarizes the oxygen atoms to give them negative hyperfine couplings. The average value of the HF coupling constant is calculated to be rather large, 506 MHz, in qualitative agreement with the observed magnitude [21] of 352.70 MHz for the OBC defect. PL from the $^2A'$ is expected at an energy of 1.90 eV [33].

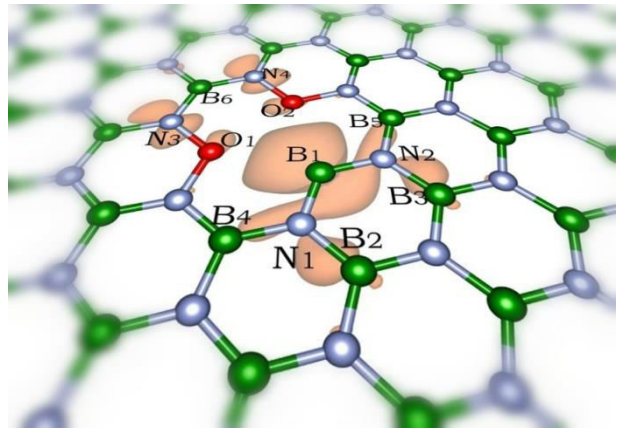

FIG.3. Geometrical structure and isosurface of the calculated spin density for $V_NO_{2B}$ centre in h-BN shown from (001) plane at isovalue 0.001/°Å$^3$. Spin density is concentrated on the labelled atoms, providing significant hyperfine couplings.

### C. Anti-site defect $V_NN_B$

Nitrogen vacancies coupled with nitrogen substitution for boron on adjacent "anti" atomic sites have been proposed as a likely source of the observed intense room temperature single photon emission from 2-D h-BN nano-flakes [7,8]. Recently, very interesting applications of spin mechanics involving cooling of a mechanical resonator by coupling spin-qubits with the $V_NN_B$ defect in h-BN have been proposed [54]. While no experimental EPR data is currently available for this defect, we simulate its properties so as to aid subsequent spectral assignment.

The calculated geometrical structure and spin density are shown in Fig. 4. The structure has $C_{2v}$ symmetry with a $^2B_1$ π-type ground state, the unpaired electron density is localized mainly on the $p_z$ orbital of the $N_1$ atom. The ground state electronic configuration of the orbitals within the h-BN band gap is $(a_1)^2(1b_1)^1(2b_1)^0$ and the PL state also is predicted to have $^2B_1$ symmetry with $(a_1)^2(1b_1)^0(2b_1)^1$ configuration, as has been previously proposed [54]. PL is thus predicted to be polarized in-plane along the $C_{2v}$ axis, with calculated adiabatic transition energy 2.12 eV [33]. This is in agreement with observations and previous calculations [7].

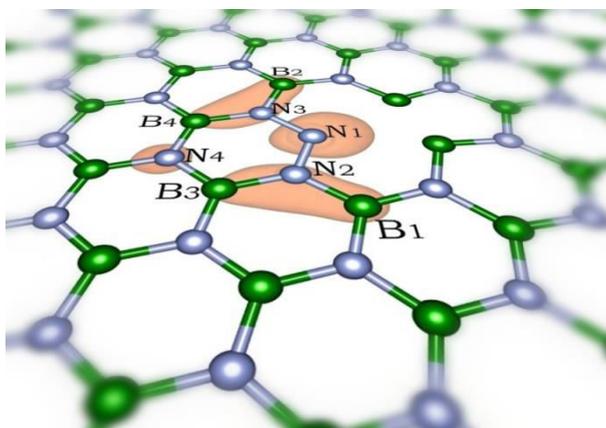

FIG.4. Geometrical structure and isosurface of the calculated spin density for $V_N N_B$ centre in h-BN shown from (001) plane at isovalue 0.001/°Å$^3$. Spin density is concentrated on the labelled atoms, providing significant hyperfine couplings

### D. Carbon and silicon related centres at nitrogen and boron vacancies $V_N C_B$, $V_B C_N$, $V_B C_N Si_N$ and $V_N C_B Si_B$.

While $V_N N_B$ have been directly observed [7], many related defects have been postulated [33] and here we consider the properties of some feasible defect sites. These sites are: $V_N C_B$ (nitrogen vacancy with one of the surrounding borons replaced with a carbon), $V_B C_N$ (boron vacancy with one of the surrounding nitrogen replaced with a carbon), $V_N C_B Si_B$ (nitrogen vacancy with one of surrounding boron replaced with carbon and another replaced with silicon) and $V_B C_N Si_N$ (boron vacancy with one of the surrounding nitrogen replaced with carbon and another replaced with silicon). For these, optimized structures and ground-state spin densities are shown in Fig. 5. Calculated defect bond lengths are provided in Supporting Data[47] Table S.II and indicate partial stabilization of the dangling bonds present in each defect. Calculated adiabatic transition energy, ground-state and excited state electronic structures are listed in Table I.

Of particular interest is the structure of $V_NC_B$ which has been previously predicted to be triplet in character [55]. However, we find the ground-state geometrical structure to be of $C_{2v}$ symmetry and $^1A_1$ in character. The properties of this defect have been the subject of recent comprehensive calculations comparing the results of DFT methods including the ones employed here to both *ab initio* calculations and various first-principles and empirical DFT calculations using model compounds to mimic the $V_NC_B$ defect [32]. These calculations indicate that the presently used method significantly *underestimates* the stability of the $^1A_1$ compared to the lowest-energy triplet state, $^3B_1$. Hence it is clear that the actual ground state should indeed be $^1A_1$. The adiabatic energy difference between these states evaluated by DFT is 0.34 eV, but adding the correction terms found for the model compound would suggest an energy difference more like 0.74 eV. However, anticipating the possibility that a triplet state is one day observed for $V_NC_B$, results for the $^3B_1$ state are also reported in Fig. 5 and Table I.

For $V_BC_N$, the ground state is calculated to be of triplet character, whereas for $V_NC_BSi_B$ and $V_BC_NSi_N$, the ground state is calculated to be doublet (see Table I). The calculated HF coupling constants for each (as well as the $^3B_1$ excited state of $V_NC_B$) are given in Table I and Supporting Data[47] Table S.II. From Fig. 5, HF constants are dominated by the spin densities which for $V_NC_B$ are found to localize on the π orbitals of the carbon and neighbouring atoms. For $V_BC_N$, the spin density is located mostly on carbon and nitrogen π orbitals, becoming highly delocalized for $V_BC_NSi_B$. Only for $V_BC_NSi_N$, does the spin density reside on σ orbitals, these showing some delocalization.

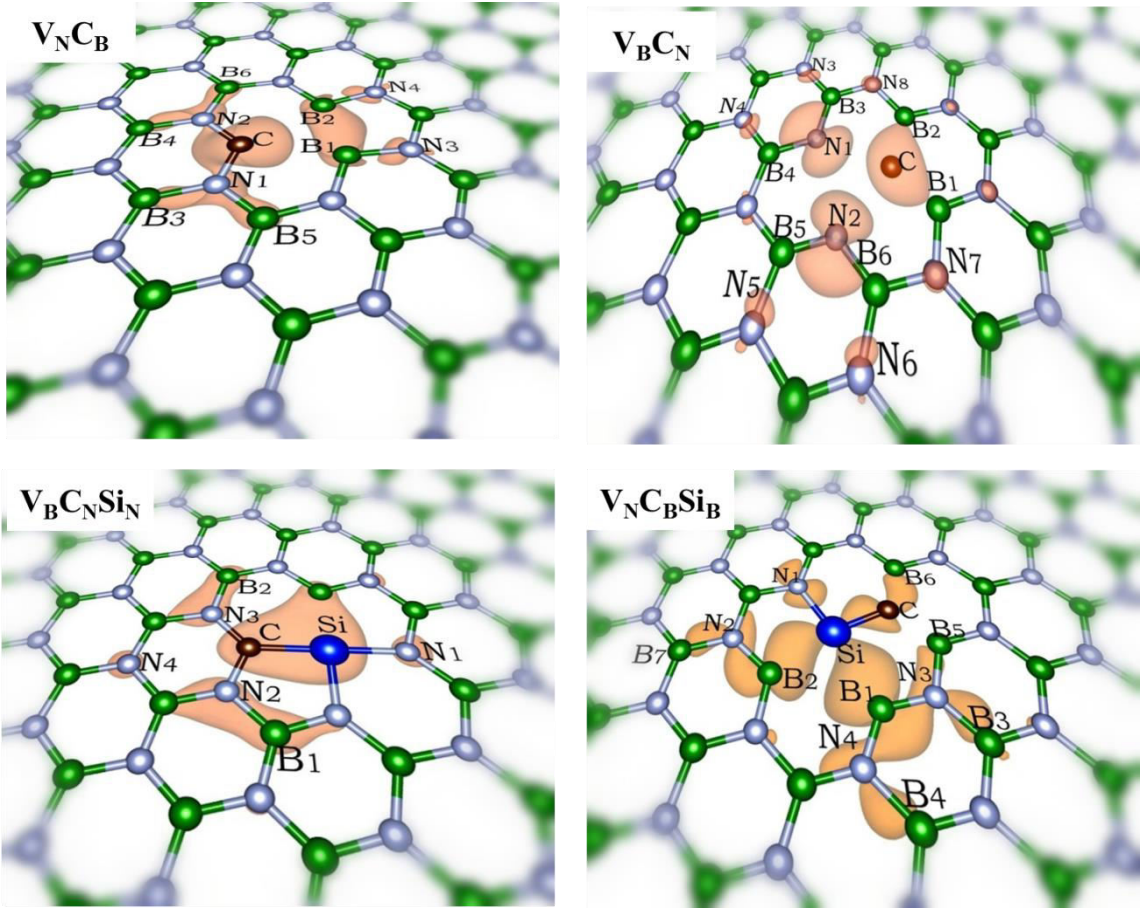

FIG.5. Geometrical structure and isosurface of the calculated spin density for $V_NC_B$ (isovalue 0.003/°Å$^3$), $V_BC_N$ (isovalue 0.004/°Å$^3$), $V_BC_NSi_N$ (isovalue 0.001/°Å$^3$) and $V_NC_BSi_B$ (isovalue 0.001/°Å$^3$) centres in h-BN shown from (001) plane. Spin density is concentrated on the labelled atoms, providing significant hyperfine couplings.

### 1. Possible photoluminescence arising from $V_NC_B$.

In h-BN, a narrow emission band has been observed with a zero phonon line (ZPL) transition at 1.95 eV [7,48,49] but the origin of this emission is unclear. Based on a broad examination of possible defect sites in h-BN using DFT with the PBE functional, we have reasoned that $V_NC_B$ forms a likely candidate as its origin [33]. Here we consider this possibility in greater depth, performing calculations using the more advanced HSE06 functional. Also, we have previously performed *ab initio* calculations using the CCSD(T) [56,57], EOMCCSD [58,59], and CASPT2[60] and MRCI[61] methods to calibrate HSE06 calculated energies for the states of $V_NC_B$ [32] and here in utilize the results to make realistic predictions of photoluminescence energies.

The electronic structure of $V_NC_B$ has been discussed in detail elsewhere [32,55], with the effects seen being generically characteristic of most h-BN defects. Basically at the defect site one σ and one π orbital on the two defect boron atoms and the carbon atom are left with dangling bonds. For the case of $V_NC_B$, four electrons need be distributed in these orbitals. The three σ atomic defect orbitals combine to make molecular orbitals depicted in Fig. 6, one of an apparently "bonding" nature (named $1a_1$), one of a "non-bonding" nature (named $2a_1$), and one of an "antibonding" nature (named $b_2$). Similarly, the three π orbitals combine to make analogous orbitals named $1b_1$, $2b_1$, and $a_2$, respectively. The "bonding" orbitals have the shape of bonding orbitals found in say 3-centre 2-electron bonds but the interatomic distances are so large that in reality no bond exists, and it is this feature that DFT methods find difficult to accurately model [32,62]. We find that $1a_1$ lies below the VB and so is always doubly occupied, $2a_1$, $1b_1$ and then $2b_1$ fall in the VB-CB gap of h-BN, while $a_2$ and $b_2$ fall inside the CB. Transitions amongst these 6 orbitals dominate the spectroscopy of the defect, with transitions involving orbitals localized within the h-BN valence-conduction band gap being the most likely to produce sharp absorption and emission spectra. Visualizations of wave functions corresponding to key Molecular Orbitals (MOs) are shown in Fig. S.1 (mid band-gap orbitals for the 2-D layer) and elsewhere [32] (all defect orbitals for a model compound).

A search of 18 excited states of $V_NC_B$ [32] identified the only ones likely to contribute to the observed 1.95 eV photoluminescence to be $(1)^1B_1$ and $(2)^1A_1$ if the luminescence occurs within the singlet manifold, and $(1)^3B_1$ and $(2)^3B_1$ if intersystem crossing leads to population of the triplet-state manifold. These states and their properties are sketched in Fig. 6(b) and are interpreted in terms of the orbital energies for the $(1)^1A_1$ ground state shown in Fig. 6(a). The energies shown in this figure include those calculated using DFT as well as those corrected using *ab initio* calculated corrections [32]. The lowest-energy adiabatic transition within the singlet manifold is predicted to be $(1)^1B_1 \rightarrow (1)^1A_1$ at 2.08eV, close to the observed value, while the lowest-energy transition within the triplet manifold is predicted to be about the same, $(2)^3B_1 \rightarrow (1)^3B_1$ at 1.58 eV. The singlet-manifold transition is allowed with an oscillator strength calculated by TDDFT for a model compound to be 0.0002 [32]. By symmetry this transition will have its dipole oriented perpendicular to the plane of the h-BN layer and emission would therefore be in the plane. There is also a singlet manifold transition with its dipole in the plane at 2.46 eV involving the doubly excited $(2)^1A_1$ state. While its dipole strength will be very small, emission with a reasonable lifetime could result following intersystem crossing to $(1)^3B_1$ state. While the best-estimate of the state energies after correction has this process endothermic by 0.16 eV, it is feasible that the reaction is exothermic instead, making it also feasible that this transition produces the observed photoemission.

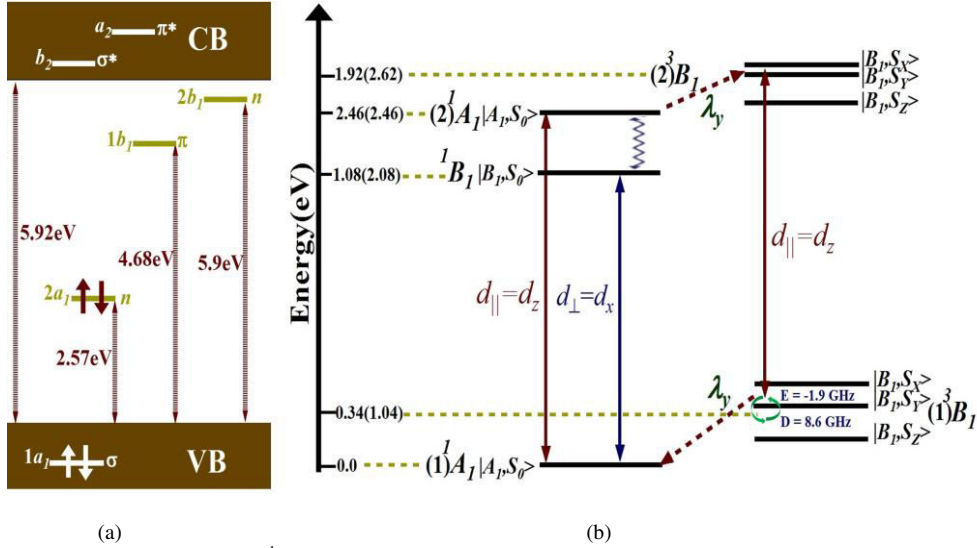

(a)             (b)

FIG.6. (a) Key DFT orbitals from the $(1)^1A_1$ closed-shell ground-state electronic structure of $V_NC_B$. The states are labelled according to the symmetry of Irreducible representation as per $C_{2v}$ point group. x(y,z)-axis are perpendicular(in the) to the plane of defect. (b) HSE06 adiabatic energies of low lying states of $V_NC_B$ as calculated by DFT, with, in (), these energies corrected according to *ab initio* CCSD(T), EOMCCSD, and CASPT2 calculations for a model compound [32]. Allowed transition polarizations *d*, spin-orbit couplings λ driving non-radiative transitions and zero-field splittings are also indicated.

### 2. Spin-orbit coupling and zero-field splitting in $V_NC_B$ and $V_BC_N$ defects.

Since spin-orbit coupling can mix the triplet and singlet spin states, generating intersystem crossings, while spin-spin interaction lifts the degeneracy of spin-multiplets, we briefly study the effect of these two interactions for $V_NC_B$ and $V_BC_N$ defects. $V_BC_N$ is chosen as it is the only defect predicted to have a triplet ground state, making it feasible for use in quantum spin devices. The electronic orbital energies of this defect are shown in Fig. 7 and are analogous to those shown in Fig. 6 for $V_NC_B$, except that the involved σ orbitals are now ordered $1a_1 < b_2 < 2a_1$ while the π orbitals are ordered $1b_1 < a_2 < 2b_1$. Its ground state is $^3B_2$ with the lowest-energy singlet state calculated to be $^1A_1$ at an adiabatic transition energy of 1.23 eV as calculated by DFT, changing to 0.53 eV applying likely corrections based on the *ab initio* calculations for $V_NC_B$. This energy difference is sufficiently high such that the prediction of a triplet ground state is likely to be robust. For $V_BC_N$, three defect orbitals (see Supporting Data Fig. S1[47]) fall within the VB of h-BN, making them doubly occupied, while the remaining three orbitals fall in the bad gap between VB and CB and are occupied by two electrons. We also consider spin-orbit coupling in $V_NC_B$ owing to its somewhat analogous electronic structure and its low-lying triplet state that could be made accessible.

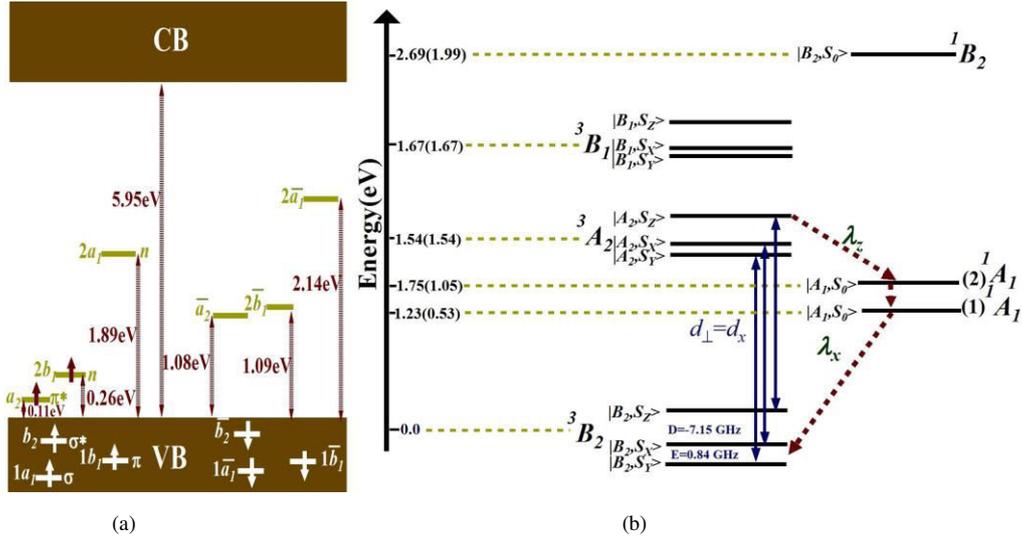

FIG.7. (a) Key DFT orbitals from the $(1)^3B_2$ ground-state electronic structure of $V_BC_N$ (the axis conventions are same as for $V_NC_B$). (b) HSE06 adiabatic energies of low lying states of $V_BC_N$ as calculated by DFT, with, in (), these energies corrected according to *ab initio* CCSD(T), EOMCCSD, and CASPT2 calculations for a model compound of $V_NC_B$ [32]. Allowed transition polarizations $d$, spin-orbit couplings λ driving non-radiative transitions, and zero-field splittings are also indicated.

The nature of the spin-orbit interactions [63], zero-field splitting, and allowed intersystem crossing and optical transitions are complex; these are discussed in detail in Supporting Data[47] Sections S2-S4, respectively. A summary of the results is presented in Figs. 6 and 7 highlighting the allowed in-plane and out of plane transitions, Corresponding HSE06 adiabatic energies of these transitions and the possible paths of relaxation to the ground state. The key conclusion reached is that through appropriate optical pumping and relaxation cycles, ground-triplet-state spin polarization can be achieved for $V_BC_N$ defect in h-BN, making it of possible use in quantum information devices [64,65]. Thus we discuss in Supporting Data[47] Sections S2-S4 how long-lived quantum memory in h-BN can be achieved for $V_NC_B$ owing to the lifetime differences of first and second order transitions from different triplet sub-states to the singlet ground state. The result for $V_BC_N$ is most significant as for this defect the triplet state is predicted to be the ground-state of the system. Thus in subsequent optical cycles ground state spin-polarization can be achieved for $V_BC_N$ as inter-system crossing results in preferential filling of $m_s = \pm 1$ spin sub-levels of triplet ground state.

## V. CONCLUSIONS

The properties of 9 possible defect centres in h-BN were examined, focusing on EPR properties such as the HF tensor of the ground state. This work is then related to prospective photoluminescence properties and assignments presented for a range of observed data concerning h-BN defects. Assignment is made of the observed [24] EPR signal at 22.43 MHz to $V_N$, the observed [23] signal at 20.83 MHz to $C_N$, the observed [19] signal at 352.70 MHz to $V_N O_{2B}$, and also tentative assignment of the observed photoemission 0-0 transition at 1.95 eV to $V_N C_B$.

Of all the defects examined, only $V_B C_N$ was predicted to have a triplet ground state. We show that the available combination of excited-state energetics, spin-orbit coupling and zero-field splitting parameters leads to a scenario in which ground-state spin polarization and long-lived quantum memory in h-BN can be achieved for $V_B C_N$ and $V_N C_B$ respectively, by optical means, making these defects of interest for use in quantum computation.

## ACKNOWLEDGMENTS

This work was supported by resources provided by the National Computational Infrastructure (NCI), and Pawsey Supercomputing Centre with funding from the Australian Government and the Government of Western Australia. AS acknowledges receipt of an Australian Postgraduate Award funded by ARC DP 150103317. Funding is also acknowledged from ARC DP 160101301 and Chinese NSF grant #1167040630.

# Defect States in Hexagonal Boron Nitride: Assignments of Observed Properties and Prediction of Properties relevant to Quantum Computation.

A. Sajid,[1,2] Jeffrey R. Reimers,[1,3] and Michael J. Ford[1*]

[1]School of Mathematical and Physical Sciences, University of Technology Sydney, Ultimo, New South Wales 2007, Australia
[2]Department of Physics, GC University Faisalabad, Allama Iqbal Road, 38000 Faisalabad, Pakistan
[3]International Centre for Quantum and Molecular Structures and School of Physics, Shanghai University, Shanghai 200444, China

*Mike.Ford@uts.edu.au

**EPAPS Supporting Data**

## S1. Supporting figures and tables

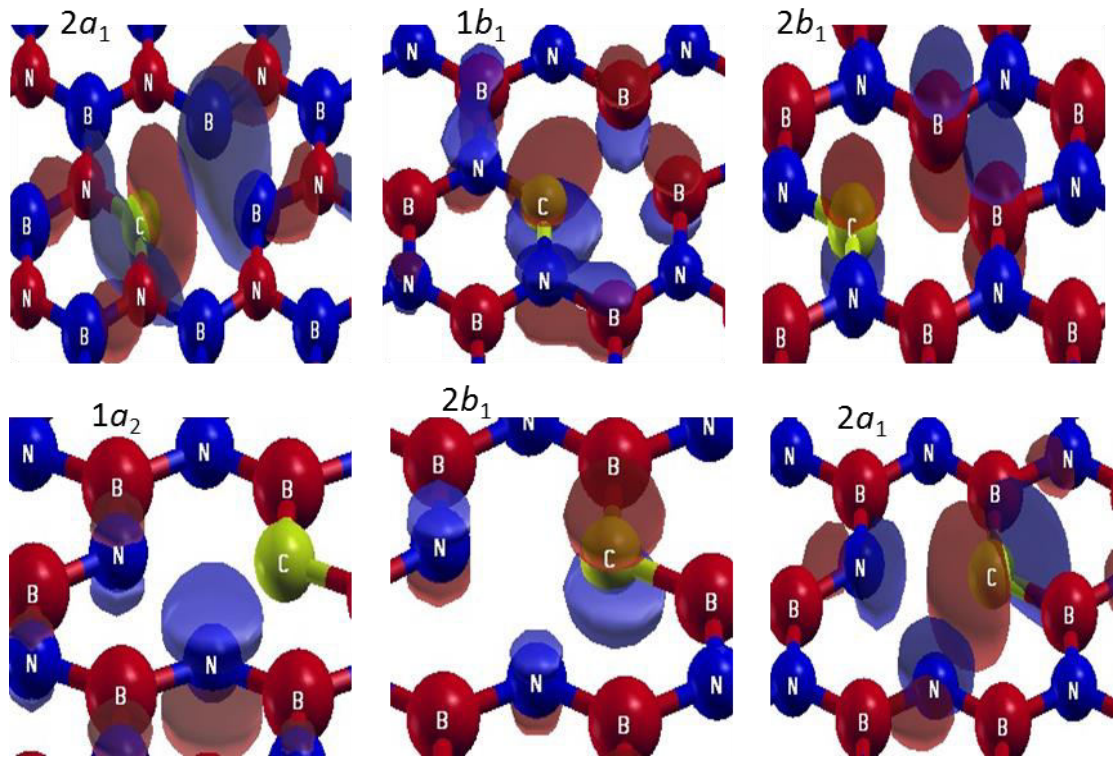

**FIG. S.1.** Orbitals localized within the h-BN valence-conduction band gap from the periodic solids of: top row: $V_NC_B$, and bottom row- $V_BC_N$.

**Table S.I** The calculated principle values of HF tensor on nearby atoms at distance $d$ from the defect centre, in MHz, for various ground-state (GS) defects in h-BN, with also the average principle values of the HF coupling tensor $(A_{xx}+A_{yy}+A_{zz})/3$. Key interatomic distances $d$ are also listed.

| Defect | Symm. | Atoms | $d$ (Å) | $A_{xx}$ | $A_{yy}$ | $A_{zz}$ | $(A_{xx}+A_{yy}+A_{zz})/3$ |
|---|---|---|---|---|---|---|---|
| $V_N$ | $D_{3h}$ | **$B_1$-$B_3$** | 1.27 | 19 | 19 | 57 | 32(22)[b] |
| | | $B_4$-$B_6$ | 2.88 | 6 | 5 | 14 | 8 |
| | | $N_1$-$N_6$ | 2.42 | -2 | -1 | -3 | -2 |
| $V_N^{-1}$ | $D_{3h}$ | **$B_1$-$B_3$** | 1.13 | 0 | 0 | 0 | 0[c] |
| $C_N$ | $D_{3h}$ | C | 0 | 40 | 40 | 208 | 96 |
| | | **$B_1$-$B_3$** | 1.49 | -16 | -14 | -22 | -17(-18)[d] |
| | | $B_4$-$B_9$ | 2.89 | -2 | -1 | -2 | -2 |
| | | $N_1$-$N_6$ | 2.53 | 1 | 0 | 5 | 2 |
| $V_NO_{2B}$ | $C_s$ | **$B_1$** | 1.41 | 529 | 519 | 615 | 554(506)[c] |
| | | $B_2$-$B_3$ | 3.77 | 21 | 21 | 27 | 23 |
| | | $B_4$-$B_5$ | 2.86 | 12 | 12 | 16 | 13 |
| | | $B_6$ | 2.90 | -3 | -2 | -5 | -3 |
| | | $N_1$-$N_2$ | 2.42 | 22 | 21 | 31 | 25 |
| | | $N_3$-$N_4$ | 2.61 | 4 | 4 | 5 | 4 |
| | | $O_1$-$O_2$ | 1.41 | -215 | -209 | -215 | -213 |
| $V_NN_B$ | $C_{2v}$ | $N_1$ | 1.37 | 31 | 29 | 139 | 66 |
| | | $N_2$-$N_3$ | 2.40 | -2 | -1 | 12 | 3 |
| | | $N_4$ | 4.25 | 1 | 1 | 5 | 2 |
| | | $B_1$-$B_2$ | 1.37 | 4 | 0 | 8 | 4 |
| | | $B_3$-$B_4$ | 3.73 | 1 | 0 | -3 | -1 |
| $^3V_NC_B$[e] | $C_{2v}$ | C | 1.35 | 497 | 418 | 503 | 473 |
| | | $B_1$-$B_2$ | 1.35 | 70 | 64 | 76 | 70 |
| | | $B_3$-$B_4$ | 3.70 | 14 | 12 | 17 | 14 |
| | | $B_5$-$B_6$ | 2.73 | 8 | 4 | 11 | 8 |
| | | $N_1$-$N_2$ | 2.36 | 7 | 5 | 10 | 7 |
| | | $N_3$-$N_4$ | 2.62 | 7 | 7 | 10 | 8 |
| $V_BC_N$ | $C_{2v}$ | C | 1.37 | -31 | 6 | 109 | 28 |
| | | $B_1$-$B_2$ | 2.58 | -5 | -5 | -8 | -6 |
| | | $B_3$-$B_4$ | 2.54 | -3 | -1 | 6 | 1 |
| | | $N_1$-$N_2$ | 1.37 | 54 | 52 | 93 | 66 |
| | | $N_3$-$N_6$ | 3.64 | 4 | 4 | 6 | 5 |
| | | $N_7$-$N_8$ | 3.04 | 1 | 0 | 3 | 1 |
| $V_BC_NSi_N$ | $C_s$ | Si | 0 | 90 | 67 | 96 | 84 |
| | | C | 2.77 | 32 | 30 | 47 | 36 |
| | | $B_1$ | 2.23 | 311 | 304 | 398 | 338 |
| | | $B_2$ | 2.08 | 166 | 164 | 206 | 179 |
| | | $B_3$ | 4.27 | 17 | 17 | 24 | 19 |
| | | $B_4$ | 4.64 | 14 | 14 | 18 | 15 |
| | | $B_5$ | 2.69 | 13 | 12 | 17 | 14 |
| | | $B_6$ | 2.89 | 12 | 11 | 16 | 13 |
| | | $B_7$ | 4.14 | 12 | 11 | 15 | 13 |
| | | $N_1$ | 1.74 | 19 | 19 | 26 | 21 |
| | | $N_2$ | 2.73 | 14 | 13 | 20 | 16 |
| | | $N_3$ | 2.87 | 13 | 11 | 20 | 15 |
| | | $N_4$ | 3.37 | 9 | 8 | 13 | 10 |
| $V_NC_BSi_B$ | $C_s$ | Si | 0 | -28 | -26 | -145 | -66 |
| | | C | 1.91 | 23 | 22 | 106 | 50 |
| | | $N_1$ | 1.79 | 2 | 1 | 4 | 2 |
| | | $N_2$-$N_3$ | 2.9 | -1 | 0 | 7 | 2 |
| | | $N_4$ | 4.79 | 1 | 1 | 3 | 2 |
| | | $B_1$ | 2.65 | 4 | 3 | 11 | 6 |
| | | $B_2$ | 3.54 | 3 | 2 | 8 | 4 |

[a] The values reported are for $B^{11}$, $N^{14}$, $C^{13}$ and $O^{17}$ isotopes, with only atoms with signal great than 2 MHz listed. Contributions from atoms interpreted as accounting for observed signals are highlighted in bold. The Fermi contact terms are listed without (with) core contribution.
[b] In ref. [24], a signal is observed of magnitude $|A_{xx}+A_{yy}+A_{zz}|/3 = 22.43\pm1.4$ MHz that is attributed to an $N_v$ centre.
[c] In Ref. [19], a signal is observed of magnitude $|A_{xx}+A_{yy}+A_{zz}|/3 = 117.06$ MHz that is attributed to an $N_V^{-1}$ TBC, as well as a signal at 352.70 MHz attributed to an oxygen-containing $V_NO_{2B}$ OBC.
[d] In Ref. [23], a signal is observed of magnitude $|A_{xx}+A_{yy}+A_{zz}|/3 = 20.83$ MHz that is attributed to the $B^{11}$ atoms in a $C_N$ TBC.
[e] The ground state is calculated to be $^1A_1$ and the results presented are for the lowest-energy triplet state, $^3B_1$.

**Table S.II** Bond lengths around $V_NN_B$, $V_NC_B$, $V_BC_N$, $N_VB_CB_{Si}$ and $B_VN_CN_{Si}$ centres.

| | | $V_NN_B$ | | $V_NC_B$ | | $V_BC_N$ | | $N_VB_CB_{Si}$ | | $B_VN_CN_{Si}$ |
|---|---|---|---|---|---|---|---|---|---|---|
| Bond | B-N | 2.57 | B-C | 2.60 | B-C | 1.55 | Si-N | 1.82 | | 1.74 |
| Length(Ang) | B-B | 1.94 | B-B | 1.92 | N-N | 2.65 | Si-B | 1.96 | | 2.08 |
| | N-N | 1.36 | C-N | 1.35 | C-N | 2.38 | Si-c | 1.91 | | 1.72 |

## S2. The spin-orbit Hamiltonian

The spin-orbit interaction is expressed in terms of the Hamiltonian [63]

$$H_{so} = \sum_k \frac{1}{2} \frac{\hbar}{c^2 m_e^2} (\nabla_k V \times P_K) \cdot \left(\frac{S_K}{\hbar}\right),$$

Where, $V = e\phi$ is the nuclear potential, $m_e$ the mass of electron and $P_K(S_K)$ is the momentum (spin) of the electron. Since the potential transforms as the totally symmetric representation ($a_1$ in $C_{2v}$ symmetry), $\nabla_k V \times P_K$ transforms as a vector $V(V_x, V_y, V_z)$. Since $P_K$ also transforms as a vector, it is possible to find the irreducible representation to which orbital operator $\vec{O} = \nabla_k V \times P_K$ belong and, using group theory, predict the non-vanishing elements of spin-orbit Hamiltonian. The only non-vanishing element of $\vec{O}$ is that for which $\langle \varphi_f | \vec{O} | \varphi_i \rangle \supset A_1$. For ground and excited states of $V_N C_B$, the spatial part of the wave function transform like either $B_1$ or $A_1$. Similarly for $V_B C_N$ the spin-orbit interaction occurs between ground state triplet and first higher energy singlet with spatial symmetries of $B_2$ and $A_1$, respectively. The non-vanishing elements of $\vec{O}$ can therefore be determined and are summarized in Table S.III for both $V_N C_B$ and $V_B C_N$.

**Table S.III.** The Non-vanishing components of spin-orbit interaction for $V_N C_B$ and $V_B C_N$. The quantities in rows and columns represent the symmetry of the spatial part of wave function for ground and first excited states of $V_N C_B$ and $V_B C_N$.

| $V_N C_B$ | $\vec{O}^{A_1}$ | $A_1$ | $B_1$ | $\vec{O}^{A_2}$ | $A_1$ | $B_1$ | $\vec{O}^{B_1}$ | $A_1$ | $B_1$ | $\vec{O}^{B_2}$ | $A_1$ | $B_1$ |
|---|---|---|---|---|---|---|---|---|---|---|---|---|
| | $A_1$ | × | 0 | $A_1$ | 0 | 0 | $A_1$ | 0 | × | $A_1$ | 0 | 0 |
| | $B_1$ | 0 | × | $B_1$ | 0 | 0 | $B_1$ | × | 0 | $B_1$ | 0 | 0 |
| $V_B C_N$ | $\vec{O}^{A_1}$ | $A_1$ | $B_2$ | $\vec{O}^{A_2}$ | $A_1$ | $B_2$ | $\vec{O}^{B_1}$ | $A_1$ | $B_2$ | $\vec{O}^{B_2}$ | $A_1$ | $B_2$ |
| | $A_1$ | × | 0 | $A_1$ | 0 | 0 | $A_1$ | 0 | 0 | $A_1$ | 0 | × |
| | $B_2$ | 0 | × | $B_2$ | 0 | 0 | $B_2$ | 0 | 0 | $B_2$ | × | 0 |

Since angular momentum transforms like an axial vector and has no $A_1$ component, we can conclude from the above table that the only non-vanishing component of $\vec{O}$ for ground and first excited state of $V_N C_B (V_B C_N)$ is the y(x)-component. Therefore the simplified form of the spin-orbit Hamiltonian can be written like

$$H_{so(V_N C_B)} = \sum_k \lambda_y S_K^y l_K^y$$

$$H_{so(V_B C_N)} = \sum_k \lambda_x S_K^x l_K^x$$

Where $\lambda_y(\lambda_x)$ is the strength of interaction. By the same argument as stated above spin mixing of the triplet and singlet (2) $^3B_1 \Leftrightarrow$ (2) $^1A_1$ ($^3A_2 \Leftrightarrow$ (1) $^1A_1$) for $V_NC_B(V_BC_N)$ is caused by $\lambda_y(\lambda_z)$ component of spin-orbit interactions as shown in Fig.6(b)(Fig.7(b)).

## S3. Spin-Spin interactions

In general a $S = 1$ electron spin system is described by a spin Hamiltonian of the following form: $H = g_e\beta \boldsymbol{B}\hat{S} + \hat{S}D\hat{S}$. Here $g_e$ is the electronic g-factor ($g_e = 2.0028 \pm 0.0003$); $\boldsymbol{B}$ is the external magnetic field and $D$ is the zero field splitting tensor. This tensor comprises the anisotropic dipolar interaction of the two electron spins forming the triplet state averaged over their wave function. This tensor is traceless and thus characterized by two parameters, $D$ and $E$ i.e. the axial and rhombic zero-field splitting parameters, respectively. This spin-spin interaction arising due to non-spherical shape of molecular orbitals lifts the degeneracy of multiplets. Therefore, we have calculated the spin-spin contribution towards zero field splitting parameters ($D$ and $E$) for high spin ground states of $V_NC_B(V_BC_N)$ as shown in Fig.6(b)(Fig.7(b)). As spin-orbit coupling links those singlets and triplets states which possess same total wave function symmetry. Therefore a large value of D for $V_NC_B$ ($V_BC_N$) means that spin selective intersystem crossing (1) $^3B_1 \Leftrightarrow$ (1) $^1A_1$ ($^3B_2 \Leftrightarrow$ (1) $^1A_1$) can preferentially fill state of $m_s = 0$ ($m_s = \pm 1$) spin projection. Also, the first order spin-orbit interaction would link states of $m_s = \pm 1$ ($m_s = 0$) spin projection of (2) $^3B_1$ ($^3A_2$) with singlets (2) $^1A_1$ ((1) $^1A_1$) for $V_NC_B(V_BC_N)$ as shown in Fig.6(b)(Fig.7(b)). Based on these arguments we discuss the possibility of achieving spin dependent photo luminescence and ground state spin-polarization for $V_NC_B(V_BC_N)$.

## S4. Dipole Allowed Transitions and Ground Triplet-State Spin Polarization

In this section, we first study the effect of dipole allowed spin-preserved transitions. Such transitions may happen via dipole interaction. $H_{di} = \sum_j \sum_\alpha d_j^\alpha E_j^\alpha$, where $d = e(x,y,z)$ is the dipole moment of the electron and $E$ is the electric field vector. These allowed transitions are induced either by the axial component of dipole moment $d_\parallel = d_z, d_y$ or out of plane component i.e. $d_\perp = d_x$. For the $V_NC_B$ ground(excited) state triplet (1) $^3B_1$((2) $^3B_1$) is split by virtue of spin-spin zero field splitting and transformed into sub-states $|B_1, S_Z\rangle$ with symmetry $B_2$ and $|B_1, S_Y\rangle, |B_1, S_X\rangle$ with symmetries $A_1$ and $A_2'$ respectively. For $V_BC_N$ the ground (excited) state triplet $^3B_2$ ($^3A_2$) is transformed into sub-states $|B_2, S_Z\rangle(|A_2, S_Z\rangle)$ with symmetry $B_1(A_1)$ and sub-states $|B_2, S_Y\rangle(|A_2, S_Y\rangle), |B_2, S_X\rangle(|A_2, S_Y\rangle)$ with symmetries $A_2(B_2)$ and $A_1(B_1)$ respectively. For $V_NC_B$ ($V_BC_N$) the spin preserved transition (1) $^3B_1 \Leftrightarrow$ (2) $^3B_1$ ($^3B_2 \Leftrightarrow {}^3A_2$) between respective magnetic sublevels occurs for the in-plane i.e. $d_\parallel = d_z$ (out of plane i.e. $d_\perp = d_x$) component of dipole moment. For $V_NC_B$ after excitation from singlet ground state((1) $^1A_1$) to the phonon sideband of the optically allowed (2) $^1A_1$ state, system can relax directly to (1) $^1A_1$ or vibronically to $^1B_1$ and then to (1) $^1A_1$ with an emission of photon. Alternatively the inter-system crossing due to spin-orbit mixing ($\lambda_y$) as a first order process(endothermic) can take place resulting in relaxation of (2) $^1A_1$ population to (2) $^3B_1$. The transition to $m_s = \pm 1$ spin sub-states of (2) $^3B_1$ is a spin selective process as only y-component of spin-orbit mixing can couple these states. We note that the spin sublevels of the triplet split even at zero magnetic field that is caused by the electron spin - electron spin dipolar interaction (zero-field splitting) because of the low-symmetry crystal field. The population of (2) $^3B_1$ can decay to (1) $^3B_1$ and then inter-system crossing can result in preferential emptying of $m_s = \pm 1$ sub-states of (1) $^3B_1$ to singlet ground state. On the other hand decay of $m_s = 0$ spin sublevel of (1) $^3B_1$ to singlet ground state is a second order process. Therefore in subsequent optical cycles, ODMR contrast can be achieved by microwave excitation (green arrows), owing to the lifetime differences of first and second order transitions from the different triplet sub-states to the singlet ground state. This behaviour is similar to $N_2V$ defect in diamond [66]. Thus $V_NC_B$, can be exploited to realize a long-living quantum memory in h-BN as has been achieved for $N_2V$ defect in diamond.

This is an exciting result as in our recent DFT computational work, several potential single photons emitting defects in h-BN are studied and among them, the $V_NC_B$ defect is shown to best fit the experimental photoluminescence (PL) line shape [33]. Also, since the energies of two ZPL's for $V_NC_B$ (i.e. (1) $^1A_1 \leftrightarrow$ (2) $^1A_1$ and (1) $^1A_1 \leftrightarrow {}^1B_1$) are matched to the energies of group-1 and group-2 emitters reported in recent experimental work [8]. Therefore our results might trigger many future studies on $V_NC_B$ in h-BN for exploring its spin-physics.

For $V_BC_N$ the excitation from ground state triplet $^3B_2$ to 2$^{nd}$ excited state triplet $^3A_2$ is forbidden due to symmetry conditions. However, not only spin dependent photo-luminescence can be achieved for $V_BC_N$ for excitation to $^3A_2$ excited state but in addition ground state spin polarization can be realized as shown in Fig.7. As z-component of spin-orbit coupling would link states of $m_s = 0$ spin projection with singlet (1) $^1A_1$ and then relaxation takes place to (2) $^1A_1$, followed by relaxation (ISC) to the ($m_s = \pm 1$) sub-state of $^3B_2$. Thus we can have preferential filling of ($m_s = \pm 1$) sub-state of $^3B_2$ for excitation from $m_s = 0$ of the same in subsequent optical cycles. Therefore ground state spin polarization can be achieved for $V_BC_N$. Thus, we conclude that both $V_NC_B$ and $V_BC_N$ can potentially be exploited for spin qubit operation [64,65].